\documentclass[lettersize,journal]{IEEEtran}
\usepackage{caption}
\usepackage{amsthm}
\usepackage{amsmath,amssymb,amsfonts}
\usepackage[linesnumbered,ruled,vlined]{algorithm2e}
\usepackage{array}
\usepackage{textcomp}
\usepackage{subfig}
\usepackage{stfloats}
\usepackage{url}
\usepackage{verbatim}
\usepackage{graphicx}
\usepackage{makecell}
\usepackage{multirow}
\usepackage{xcolor}
\usepackage{cite}
\usepackage{booktabs}
\usepackage{diagbox}
\newcommand{\figsubref}[2]{Fig.~\ref{#1}\subref{#2}}

\newtheorem{proposition}{Proposition}

\begin{document}

\title{Implicit Neural Representation of Beamforming for Continuous Aperture Array Systems}


\author{Shiyong~Chen,~\IEEEmembership{Student Member,~IEEE,} Jia~Guo,~\IEEEmembership{Member,~IEEE,}
        and Shengqian~Han,~\IEEEmembership{Senior Member,~IEEE}%

\thanks{Shiyong Chen is with the School of Electronics and Information Engineering, Beihang University, Beijing 100191, China (email: shiyongchen@buaa.edu.cn).}
\thanks{Jia Guo is with the School of Electronic Engineering and Computer
Science, Queen Mary University of London, London E1 4NS, U.K. (email: jia.guo@qmul.ac.uk).}
\thanks{Shengqian Han is with the School of Electronics and Information Engineering, Beihang University, Beijing 100191, China (email: sqhan@buaa.edu.cn).}
\thanks{This work was supported by the National Natural Science Foundation of China (NSFC) under Grant 62401033.}
}



\maketitle

\begin{abstract}
In this paper, a learning-based approach is proposed for optimizing downlink beamforming in multiple-input multiple-output (MIMO) systems that employ continuous aperture arrays (CAPAs) at both the base station (BS) and the user. Beamforming in such systems is a spatially continuous function that maps a coordinate on the CAPA to a corresponding beamforming weight. We first propose an implicit neural representation (INR), termed BeaINR, to parameterize this function directly. Further, noting that the optimal beamforming function can be expressed as a weighted integral of the channel response function, we propose a second INR, CoefINR, to represent the weighting coefficient function, which indirectly optimizes the beamforming function. Simulation results show that the proposed INR-based methods achieve comparable or higher spectral efficiency (SE) than the considered baselines, while requiring substantially lower inference latency. Moreover, CoefINR reduces training complexity and improves frequency generalizability relative to BeaINR by leveraging the optimal beamforming structure.
\end{abstract}


\section{Introduction}
Continuous aperture array (CAPA) systems are recognized as a candidate technique for future wireless communications~\cite{CAPA_based}. Compared with traditional spatially discrete antenna array (SPDA) systems, CAPA systems offer greater spatial degrees of freedom and beamforming flexibility~\cite{Near_field}. A CAPA system can be regarded as an extreme SPDA system where the number of antennas approaches infinity. In this scenario, the discrete beamforming design for an SPDA system becomes a problem of designing a continuous current distribution for CAPA systems. This transition from a discrete to a continuous domain results in non-convex infinite-dimensional functional optimization problems, which are generally intractable using traditional convex optimization methods. 

Several methods have been proposed to optimize beamforming for CAPA systems. A common approach is to approximate the continuous channel and beamforming functions using a finite set of orthogonal Fourier basis functions, converting the original functional optimization into a finite-dimensional parameter optimization~\cite{Wavenumber, Pattern_Division, On_the_SE}. In~\cite{Wavenumber}, this Fourier-based technique was applied to single-user multi-stream CAPA systems and later extended to multi-user uplink and downlink scenarios~\cite{Pattern_Division, On_the_SE}. However, this method introduces approximation errors that compromise optimality. Additionally, the required number of basis functions increases rapidly with the carrier frequency and CAPA size, leading to high-dimensional optimization and substantial computational overhead in large-scale CAPA systems~\cite{CAPA}.

To eliminate the approximation errors, direct functional optimization methods leveraging the calculus of variations (CoV) have been explored for the design of CAPA beamforming~\cite {Multi_Group_Multicast, Beamforming_Design}. In ~\cite{Multi_Group_Multicast}, an iterative CoV-based block coordinate descent algorithm was proposed for downlink multicast scenarios, where the CAPA-based base station (BS) serves multiple single-antenna users. For the scenarios where both transmitters and receivers are equipped with CAPAs, an iterative weighted minimum mean-squared error (WMMSE) algorithm was derived based on CoV~\cite{Beamforming_Design}. Nevertheless, the repeated integrals and sequential updates involved in such iterative methods incur high inference time, hindering real-time~implementations.  

To reduce computational complexity and enable real-time implementations, deep neural networks (DNNs) have been applied in beamforming design for SPDA systems, where finite-dimensional beamforming matrices are directly learned from channel matrices~\cite{Recursive_GNNs,Gradient_GNN}. However, the optimization of the beamforming in CAPA systems involves continuous channel response and beamforming functions, which are infinite-dimensional and thus incompatible with standard DNN input-output formats. To address this challenge, recent work~\cite{Multi_User_CAPA} found finite-dimensional representations of the continuous beamforming function for the scenario where a CAPA-based BS serves multiple single-antenna users. Specifically, the optimal beamforming function can be represented as a weighted sum of channel response functions, enabling indirect optimization through learning the finite-dimensional weighting coefficient vector. Yet, this method does not apply to systems where both transmitter and receiver are equipped with CAPAs, as the weighting coefficient vectors become continuous weighting coefficient functions in this case, again incompatible with DNN input-output formats. 

Implicit neural representations (INRs) are a class of neural networks that represent continuous functions implicitly by mapping spatial or temporal coordinates to corresponding function values~\cite{Implicit_Neural}. In this paper, we study the optimization of downlink beamforming in MIMO systems that employ CAPAs at both the BS and the user. We first design an INR, BeaINR, to directly learn the beamforming function. BeaINR maps spatial coordinates on the BS CAPA to beamforming weights, thereby avoiding the need to learn an infinite-dimensional input-output mapping. Furthermore, we observe that the optimal beamforming resides in the functional subspace spanned by the channel response function, which can be expressed as a weighted integral of the channel response function. Thus, we design CoefINR, an INR that represents the weighting coefficient function defined on the user CAPA, with which the beamforming can be obtained indirectly. Since the user CAPA is typically smaller than the BS CAPA, the training complexity for CoefINR is effectively reduced. Simulation results show that the proposed INR-based methods achieve comparable or higher spectral efficiency (SE) than the considered baselines with significantly reduced inference latency. In addition, CoefINR reduces training complexity and improves frequency generalizability relative to BeaINR by exploiting the optimal beamforming structure.

\section{System Model and Problem Formulation}
\begin{figure}
\centering
\includegraphics[width=0.3\textwidth]{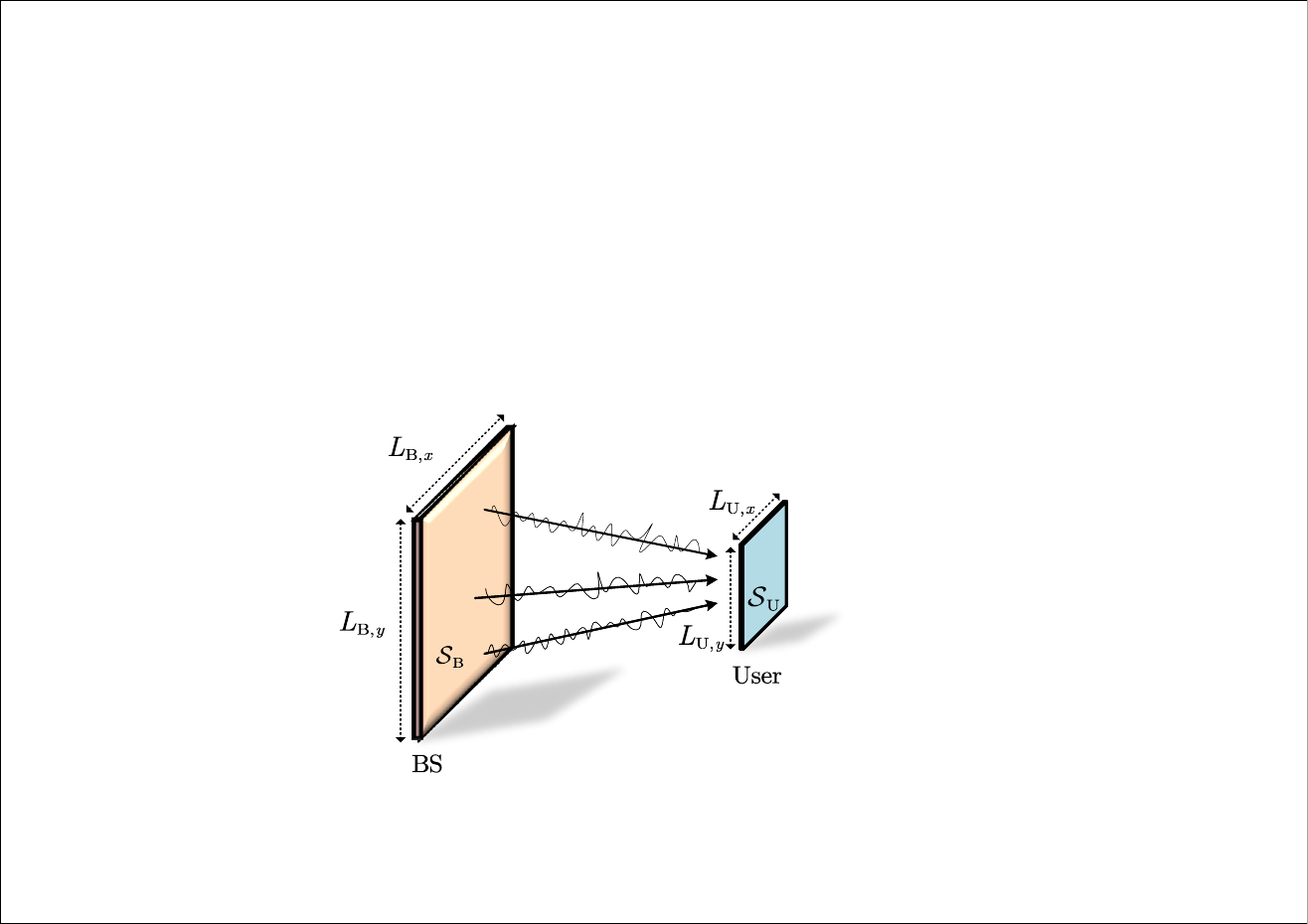}
\caption{Illustration of the downlink CAPA system. }  \label{CAPA}
\vspace{-0.3cm}
\end{figure}

Consider a downlink CAPA system where a BS serves a user, both equipped with CAPAs, as illustrated in Fig.~\ref{CAPA}. In a global coordinate system, the BS CAPA lies on the $xy$-plane, centered at the origin, with lengths $L_{\mathrm{B},x}$ and $L_{\mathrm{B},y}$ along the $x$- and $y$-axes. A point on the BS CAPA is denoted as $\mathbf{s}=[s_x, s_y, 0]^{\mathsf{T}}$, and $\mathcal{S}_\mathrm{B}$ is the set of all points on the BS CAPA. The user's position is denoted as $\mathbf{r}_o$, and the user CAPA with lengths $L_{\mathrm{U},x}$ and $L_{\mathrm{U},y}$, is centered at $\mathbf{r}_o$. To ensure polarization alignment for optimal signal reception, the user CAPA is assumed to be parallel to the BS CAPA~\cite{Beamforming_Design}.  To facilitate coordinate modeling on the user CAPA, a local coordinate system is established, where $\mathbf{r}_o$ is the origin, and the $xy$-plane aligns with the BS CAPA. In this local coordinate system, the point on the user CAPA is defined as $\bar{\mathbf{r}}=[\bar{r}_x, \bar{r}_y, 0]^{\mathsf{T}}$ and $\bar{\mathcal{S}}_\mathrm{U}$ consists of all the points in the user CAPA. Each point can be mapped to the global coordinate system via 
\begin{equation} \label{rotation equation}
    \mathbf{r} =\bar{\mathbf{r}} + \mathbf{r}_o,\quad \bar{\mathbf{r}}\in\bar{\mathcal{S}}_\mathrm{U}
\end{equation}
where $\mathbf{r}$ denotes the corresponding point in the global coordinate system and $\mathcal{S}_\mathrm{U}$ represents the set of all such points.

The BS transmits $N$ independent data streams. The signal at point $\mathbf{s} \in \mathcal{S}_\mathrm{B}$ is given by $\mathrm{J}(\mathbf{s}) = \sum_{n=1}^{N} \mathrm{w}_n(\mathbf{s}) x_n$, where $x_n$ is the data of the $n$-th stream with $\mathbb{E}\{|x_n|^2\} = 1$, and $\mathrm{w}_n(\mathbf{s})$ denotes the beamforming function for the $n$-th stream with point $\mathbf{s}$ as the input. Then the received signal at point $\mathbf{r}$ of the user CAPA can be expressed as~\cite{Multi_Group_Multicast}
\begin{equation} \label{y}
y(\mathbf{r}) = \textstyle\int_{\mathcal{S}_\mathrm{B}} h(\mathbf{r}, \mathbf{s}) \mathrm{J}(\mathbf{s}) \mathrm{d}\mathbf{s} + z(\mathbf{r}),
\end{equation}
where $h(\mathbf{r}, \mathbf{s})$ is the continuous channel response between point $\mathbf{s}$ at the BS CAPA and point $\mathbf{r}$ at the user CAPA, and $z(\mathbf{r}) \sim \mathcal{CN}(0, \sigma^2)$ denotes white thermal noise with $\sigma^2$ being noise variance.

As widely used in the literature~\cite{Beamforming_Optimization}, we consider uni-polarized CAPAs under line-of-sight conditions where both the BS and user CAPAs are polarized along the $y$-axis. The channel response function $h(\mathbf{r}, \mathbf{s})$ can be expressed as
\begin{equation} \label{Green's function}
h(\mathbf{r}, \mathbf{s})\!=\!\mathbf{u}^{\mathsf{T}}\frac{-j \eta e^{-j \frac{2\pi}{\lambda} \| \mathbf{r} - \mathbf{s} \|}}{2\lambda \| \mathbf{r} - \mathbf{s} \|} 
\left(\mathbf{I}_3\!-\!\frac{(\mathbf{r} - \mathbf{s})(\mathbf{r} - \mathbf{s})^{\mathsf{T}}}{\| \mathbf{r} - \mathbf{s} \|^2} \right)\mathbf{u},
\end{equation}
where $\mathbf{u} = [0, 1, 0]^{\mathsf{T}}$ denotes the unit polarization vector, $\eta$ is the intrinsic impedance, $\lambda$ is the signal wavelength, and $\mathbf{I}_3$ is the identity matrix of size $3\times 3$. To focus on beamforming optimization, we assume perfect channel state information at the BS. To acquire the continuous channels, existing work has proposed parametric estimation methods~\cite{Parametric_Channel_Estimation,Fourier_Plane_Wave}, which estimate a finite set of channel parameters and then reconstruct the channel response functions.

With~\eqref{y}, the beamforming optimization problem, aimed at maximizing SE, can be formulated as~\cite{Beamforming_Design}
\begin{subequations} \label{P0:Power0}
    \begin{align}
    \max_{\mathbf{w}(\mathbf{s})} \quad & \log_2 \det \Big( \mathbf{I}_N + \frac{1}{\sigma^2} \mathbf{Q} \Big) \label{eq:object} \\
    \text{s.t.} \quad 
    & \mathbf{Q} =  \textstyle\iint_{\mathcal{S}_\mathrm{U}} \mathbf{e}^{\mathsf{H}}(\mathbf{r}_1)\delta(\mathbf{r}_1-\mathbf{r}) \mathbf{e}(\mathbf{r}) \, \mathrm{d}\mathbf{r}_1\mathrm{d}\mathbf{r}, \label{eq:E1} \\
    & \mathbf{e}(\mathbf{r}) =  \textstyle\int_{\mathcal{S}_\mathrm{B}} h(\mathbf{r}, \mathbf{s}) \mathbf{w}(\mathbf{s}) \, \mathrm{d}\mathbf{s}, \label{eq:E2} \\
    &\| \mathbf{w}(\mathbf{s}) \|^2 \leq \mathrm{I}_{\max}, \forall \mathbf{s} \in{\mathcal{S}_\mathrm{B}}, \label{eq:PeakConstraint}\\
    &  \textstyle\int_{\mathcal{S}_\mathrm{B}} \| \mathbf{w}(\mathbf{s}) \|^2 \, \mathrm{d}\mathbf{s} \leq \mathrm{C}_{\max}, \label{eq:PowerConstraint}
    \end{align}
\end{subequations}
where $\mathbf{w}(\mathbf{s})=[\mathrm{w}_1(\mathbf{s}), \cdots, \mathrm{w}_N(\mathbf{s})]\in\mathbb{C}^{1\times N}$, $\mathbf{e}(\mathbf{r})\in\mathbb{C}^{1\times N}$ denotes the effective channel gain at $\mathbf{r}$, $\|\cdot\|$ is the $\ell_2$ norm, $\delta(\cdot)$ is the Dirac delta function, and $\mathrm{I}_{\max}$ and $\mathrm{C}_{\max}$ denote the peak and total transmit current budgets, respectively.

\section{INR of Beamforming Function} \label{Training Method}
In this section, we design an INR, termed BeaINR, to learn the continuous beamforming function.

The beamforming function is the solution of the functional optimization  problem~\eqref{P0:Power0}. By substituting~\eqref{rotation equation} into~\eqref{eq:E2}, then into~\eqref{eq:E1} and evaluating the integral over $\mathcal{S}_\mathrm{U}$, it follows that $\mathbf{Q}$ depends on the value of $\mathbf{r}_o$, which characterizes the relative position between the BS and the user. Therefore, the solution $\mathbf{w}(\mathbf{s})$ is also influenced by $\mathbf{r}_o$. To account for this dependency, BeaINR is designed to learn the mapping from $\mathbf{s}$ and $\mathbf{r}_o$ to the beamforming output, i.e.,
\begin{equation} \label{BeaINR}
\mathbf{w}(\mathbf{s}) = \mathcal{P}_{\theta_{\mathrm{p}}}(\mathbf{s}, \mathbf{r}_o), \quad \mathbf{s} \in \mathcal{S}_{\mathrm{B}},
\end{equation}
where $\mathcal{P}_{\theta_{\mathrm{p}}}(\cdot)$ denotes the INR-based beamforming function parameterized by $\theta_{\mathrm{p}}$.

Supervised learning for BeaINR is impractical, as learning the continuous mapping from coordinates to beamforming values requires an infinite training set. Unsupervised learning for BeaINR, which uses the objective function~\eqref{eq:object} as the loss function, is also challenging due to the intractable integrals involved in the computation of $\mathbf{Q}$. To overcome this challenge, we adopt Gauss–Legendre (GL) quadrature to approximate the integrals. Specifically, for a continuous function $f(\cdot)$ defined on $\mathcal{S}_{\mathrm{B}}$, the integral is approximated as~\cite{Beamforming_Design}
\begin{align} \label{GL of S}
 \textstyle\int_{\mathcal{S}_{\mathrm{B}}} f(\mathbf{s})\mathrm{d}\mathbf{s}
\!\approx\! \sum\limits_{m=1}^{M_{\mathrm{B,G}}} \sum\limits_{n=1}^{M_{\mathrm{B,G}}} \frac{\xi^{\mathrm{B}}_m \xi^{\mathrm{B}}_n A_{\mathrm{B}}}{4} f(\mathbf{s}_{m,n}),\quad\mathbf{s}_{m,n} \in \mathcal{S}_{\mathrm{B}},
\end{align}
where $\mathbf{s}_{m,n}\!=\!\bigl[ \frac{\phi^{\mathrm{B}}_m L_{\mathrm{B},x}}{2},\, \frac{\phi^{\mathrm{B}}_n L_{\mathrm{B},y}}{2},\, 0 \bigr]^{\mathsf{T}}$ denotes a sampling point, $A_{\mathrm{B}} =   L_{\mathrm{B},x} L_{\mathrm{B},y}$ is the CAPA area, $M_{\mathrm{B,G}}$ is the quadrature order, and $\{\phi^{\mathrm{B}}_m\}_{m=1}^{M_{\mathrm{B,G}}}$ and $\{\xi^{\mathrm{B}}_m\}_{m=1}^{M_{\mathrm{B,G}}}$ are known as the roots and weights of the Legendre polynomial, respectively.\footnote{The approximation in~\eqref{GL of S} is accurate even with a small value of $M_{\mathrm{B,G}}$. For instance, for the function $\mathbf{w}(\mathbf{s})$ learned by the BeaINR, our numerical results indicate that when $M_{\mathrm{B,G}}\geq4$, the GL quadrature yields a relative approximation error below $0.05\%$.}

With~\eqref{GL of S}, we first construct a training dataset for BeaINR, where each sample consists of a fixed set of GL quadrature points and a randomly generated user position. Specifically, $M_{\mathrm{B,G}}$ GL roots $\{\phi^{\mathrm{B}}_m\}_{m=1}^{M_{\mathrm{B,G}}}$ are used to sample $M_{\mathrm{B,G}}^2$ spatial points on $\mathcal{S}_{\mathrm{B}}$, denoted as $\{\mathbf{s}_{m,n}\}_{m,n=1}^{M_{\mathrm{B,G}}}$, which are shared across all samples. For each sample, the user position is independently generated. The resulting training dataset is defined as
\begin{equation} \label{Dataset D}
\mathcal{D}_{\mathrm{B,G}} = \Big\{ \Big( \left\{ \mathbf{s}_{m,n} \right\}_{m,n=1}^{M_{\mathrm{B,G}}},\, \mathbf{r}_o^{(k)}\Big) \Big\}_{k=1}^{\left|\mathcal{D}_{\mathrm{B,G}}\right|},
\end{equation}
where $\mathbf{r}_o^{(k)}$ denotes the user position in the $k$-th sample, and $\left|\mathcal{D}_{\mathrm{B,G}}\right|$ is the total number of samples. 

The loss over $\mathcal{D}_{\mathrm{B,G}}$ is computed using GL quadrature. Specifically, for the $k$-th sample, the beamforming function is obtained as 
\begin{equation} \label{beamforming vector}
\!\mathbf{w}^{(k)}(\mathbf{s}_{m,n}) \!= \!\mathcal{P}_{\theta_{\mathrm{p}}}(\mathbf{s}_{m,n}, \mathbf{r}_o^{(k)}),\quad \mathbf{s}_{m,n}\in\mathcal{S}_{\mathrm{B}}.
\end{equation}
Then the integral in~\eqref{eq:E2} is approximated as 
\begin{equation} \label{integral of e}
 \textstyle\mathbf{e}^{(k)}(\mathbf{r}^{(k)})\! \approx \!\sum\limits_{m=1}^{M_{\mathrm{B,G}}}\! \sum\limits_{n=1}^{M_{\mathrm{B,G}}}\!\! \frac{\xi^{\mathrm{B}}_{m}\xi^{\mathrm{B}}_{n}A_{\mathrm{B}}}{4}\! \cdot\! h(\mathbf{r}^{(k)}, \mathbf{s}_{m,n}) \!\cdot\! \mathbf{w}^{(k)}(\mathbf{s}_{m,n}),
\end{equation}
where $h(\mathbf{r}^{(k)}, \mathbf{s}_{m,n})$ denotes the channel response between point $\mathbf{s}_{m,n}$ and $\mathbf{r}^{(k)}$. Similarly, the integral in~\eqref{eq:E1} is derived~as 
\begin{equation} \label{integral of Q}
\textstyle\mathbf{Q}^{(k)} \approx \sum\limits_{m=1}^{M_{\mathrm{U,G}}} \sum\limits_{n=1}^{M_{\mathrm{U,G}}} \frac{\xi^{\mathrm{U}}_{m}\xi^{\mathrm{U}}_{n}A_{\mathrm{U}}}{4} \cdot \mathbf{e}^{(k),{H}}(\mathbf{r}^{(k)}_{m,n}) \cdot \mathbf{e}^{(k)}(\mathbf{r}^{(k)}_{m,n}),
\end{equation}
where $A_{\mathrm{U}} = L_{\mathrm{U},x} L_{\mathrm{U},y}$ is the area of $\mathcal{S}_{\mathrm{U}}$, $M_{\mathrm{U,G}}$ denotes quadrature order to sample points on $\mathcal{S}_{\mathrm{U}}$, $\{\xi^{\mathrm{U}}_m\}_{m=1}^{M_{\mathrm{U,G}}}$ are the weights of the Legendre polynomial for integral approximation on $\mathcal{S}_{\mathrm{U}}$,  and $\mathbf{r}^{(k)}_{m,n}$ are the corresponding sampling points in the $k$-th sample. To obtain $\mathbf{r}^{(k)}_{m,n}$, we first apply $M_{\mathrm{U,G}}$ quadrature roots $\{\phi^{\mathrm{U}}_m\}_{m=1}^{M_{\mathrm{U,G}}}$ to generate sampling points $\{ \bar{\mathbf{r}}_{m,n} \}_{m,n=1}^{M_{\mathrm{U,G}}}$ on $\mathcal{S}_{\mathrm{U}}$ in the local coordinate system. Then, for the $k$-th sample, these points are mapped to global coordinates via $\mathbf{r}^{(k)}_{m,n} = \bar{\mathbf{r}}_{m,n} + \mathbf{r}_o^{(k)}$. Substituting $\mathbf{Q}^{(k)}$ into~\eqref{eq:object} yields the achievable SE for the $k$-th sample. The loss is defined as the negative SE, denoted by $\mathcal{L}_{\mathrm{B,G}}^{(k)}$. The average loss over dataset  $\mathcal{D}_{\mathrm{B,G}}$ is then computed as $\mathcal{L}_{\mathrm{B,G}} = \frac{1}{\left|\mathcal{D}_{\mathrm{B,G}}\right|} \sum_{k=1}^{\left|\mathcal{D}_{\mathrm{B,G}}\right|} \mathcal{L}_{\mathrm{B,G}}^{(k)}$.

Training BeaINR with fixed GL sampling points in $\mathcal{D}_{\mathrm{B,G}}$, i.e., $\{\mathbf{s}_{m,n}\}_{m,n=1}^{M_{\mathrm{B,G}}}$, may cause overfitting to these specific coordinates and limit generalization on the CAPA. To mitigate this, motivated by~\cite{NeRF}, we combine deterministic quadrature with randomized sampling for network training. Specifically, we construct an auxiliary dataset using sample-wise randomized points generated via Owen-scrambled Sobol sequence~\cite{Practical_Hash}, and train BeaINR with the combined loss from both datasets. The GL quadrature provides high-precision approximations of integrals for smooth functions, enabling fast and stable convergence with relatively few sampling points~\cite{Methods_of}. Randomized sampling introduces variability across samples, which prevents the network from memorizing fixed GL quadrature nodes and forces it to capture the global structure of the beamforming function, thereby mitigating overfitting and improving generalization across the CAPAs. Furthermore, the low-discrepancy property of the Sobol sequence ensures more uniform coverage of the CAPA compared with the standard independent and identically distributed random sampling, reducing the errors of the integral approximation and accelerating convergence~\cite{Practical_Hash}.

In the auxiliary dataset, we generate  $M_{\mathrm{B, S}}$ randomized sampling points for each sample. The generating procedure is summarized in \textbf{Algorithm 1}, where \texttt{sobolset}(2), $\texttt{scramble}(P,\texttt{\textquotesingle Owen\textquotesingle})$ and \texttt{net}($P, M_{\mathrm{B,S}}$) are built-in functions in MATLAB to construct a 2D Sobol sequence generator, add Owen scrambling to the Sobol sequence, and generate the first $M_{\mathrm{B,S}}$  sample points from the scrambled Sobol generator, respectively. The resulting sampling points $\bigl\{\!\mathbf{s}_{i}^{(k)} \!\bigr\}_{i=1}^{M_{\mathrm{B,S}}}$, along with the corresponding user position $\mathbf{r}_o^{(k)}$, form the auxiliary~dataset
\begin{equation} \label{Dataset Ds}
\mathcal{D}_{\mathrm{B,S}} = \Big\{ \Big( \big\{ \mathbf{s}_{i}^{(k)} \big\}_{i=1}^{M_{\mathrm{B,S}}},\, \mathbf{r}_o^{(k)} \Big) \Big\}_{k=1}^{\left|\mathcal{D}_{\mathrm{B,S}}\right|},
\end{equation}
where $\left|\mathcal{D}_{\mathrm{B,S}}\right|$ denotes the size of the dataset. 

\begin{table}[!t]
\centering
\begin{tabular}{p{0.95\columnwidth}}
\toprule  
\textbf{Algorithm 1:} Generation of Sample-wise Randomized Points\\
\midrule
\textbf{Input:} Number of points $M_{\mathrm{B,S}}$, aperture lengths $L_{\mathrm{B},x}$ and $L_{\mathrm{B},y}$, sample index $k$.\\
\textbf{Output:} Sampling points $\{\mathbf{s}_{i}^{(k)}\}_{i=1}^{M_{\mathrm{B,S}}}$ for the $k$-th sample, with $\mathbf{s}_i^{(k)} = [s_{i,1}^{(k)}, s_{i,2}^{(k)}, 0]^{\mathsf T}$.\\[2pt]

1:\quad Set the random seed to $k$.\\
2:\quad Construct a 2D Sobol sequence generator $P \leftarrow \texttt{sobolset}(2)$.\\
3:\quad Apply Owen scrambling to $P$: $P \leftarrow \texttt{scramble}(P,\texttt{\textquotesingle Owen\textquotesingle})$.\\
4:\quad Generate $M_{\mathrm{B,S}}$ quasi-random points on $[0,1)^2$: 
$\mathbf{U} = [\mathbf{u}_1^{\mathsf T},\dots,\mathbf{u}_{M_{\mathrm{B,S}}}^{\mathsf T}]^{\mathsf T} 
\leftarrow \texttt{net}(P, M_{\mathrm{B,S}})$, where $\mathbf{u}_i = [u_{i,1}, u_{i,2}]^{\mathsf T}$.\\

5:\quad \textbf{for} $i = 1$ to $M_{\mathrm{B,S}}$ \textbf{do}\\
6:\qquad Map to physical aperture coordinates:
$\mathbf{s}_i^{(k)} = [s_{i,1}^{(k)}, s_{i,2}^{(k)}, 0]^{\mathsf T}$, where
$s_{i,1}^{(k)} = L_{\mathrm{B},x}(u_{i,1}-0.5)$ and $s_{i,2}^{(k)} = L_{\mathrm{B},y}(u_{i,2}-0.5)$.\\
7:\quad \textbf{end for}\\

8:\quad Return $\{\mathbf{s}_{i}^{(k)}\}_{i=1}^{M_{\mathrm{B,S}}}$.\\
\bottomrule
\end{tabular}
\vspace{-0.3cm}
\end{table}

The loss of the $k$-th sample in $\mathcal{D}_{\mathrm{B,S}}$, denoted by $\mathcal{L}_{\mathrm{B,S}}^{(k)}$, is computed following the same procedure as $\mathcal{L}_{\mathrm{B,G}}^{(k)}$, but with a different integral approximation. Specifically, with the sampled points $\{\mathbf{s}_{i}^{(k)}\}_{i=1}^{M_{\mathrm{B,S}}}$, the integral of function $f(\cdot)$ defined on $\mathcal{S}_{\mathrm{B}}$ is approximated as
\begin{align} \label{OS of S}
 \textstyle\int_{\mathcal{S}_{\mathrm{B}}} f(\mathbf{s})\, \mathrm{d}\mathbf{s}\approx \sum_{i=1}^{M_{\mathrm{B,S}}} \frac{f(\mathbf{s}^{(k)}_{i})}{M_{\mathrm{B,S}}}A_{\mathrm{B}}.
\end{align}
The average loss over the dataset $\mathcal{D}_{\mathrm{B,S}}$ is given by $\mathcal{L}_{\mathrm{B,S}} = \frac{1}{\left|\mathcal{D}_{\mathrm{B,S}}\right|} \sum_{k=1}^{\left|\mathcal{D}_{\mathrm{B,S}}\right|} \mathcal{L}_{\mathrm{B,S}}^{(k)}.$
 
Finally, the overall loss for training the BeaINR is formulated as
\begin{equation}
\mathcal{L}_{\mathrm{B}} = (1-\alpha)\cdot\mathcal{L}_{\mathrm{B,G}} + \alpha \cdot \mathcal{L}_{\mathrm{B,S}},
\label{eq:total_loss}
\end{equation}
where $\alpha\in[0,1]$ is a factor controlling the importance of $\mathcal{L}_{\mathrm{B,S}}$.

\section{INR of Weighting Coefficient Function} \label{CoefINR Sec}
Prior studies have shown that in SPDA~\cite{A_universal} or CAPA systems with single-antenna users~\cite{Multi_User_CAPA}, the optimal beamforming can be represented as a linear combination of the channel matrices or channel response functions, respectively. The scenario considered in this paper can be regarded as an extreme SPDA system where both the number of transmit and receive antennas approaches infinity. Accordingly, the discrete combining coefficients in~\cite{A_universal} and~\cite{Multi_User_CAPA} generalize to continuous weighting coefficient functions. The following proposition formalizes this generalization.

\begin{proposition} \label{Proposition1}
The optimal beamforming function \( \mathbf{w}(\mathbf{s}) \) lies in the functional subspace spanned by the channel response function \( \{ h(\mathbf{r}, \mathbf{s}) \}_{\mathbf{r}\in\mathcal{S}_\mathrm{U}} \). That is, \( \mathbf{w}(\mathbf{s}) \) can be represented as
\begin{equation} \label{optimal structure}
    \mathbf{w}(\mathbf{s}) = \textstyle \int_{\mathcal{S}_{\mathrm{U}}} h(\mathbf{r}, \mathbf{s})\, \mathbf{c}(\mathbf{r}) \,\mathrm{d}\mathbf{r},
\end{equation}
where \( \mathbf{c}(\mathbf{r}) = [c_1(\mathbf{r}), \cdots, c_N(\mathbf{r})] \in \mathbb{C}^{1\times N} \) denotes a set of weighting coefficient functions defined on the user CAPA, and \( c_n(\mathbf{r}) \) corresponds to the \( n \)-th data stream.
\end{proposition}
\begin{IEEEproof}
The proof is omitted due to space constraints.
\end{IEEEproof}

Motivated by Proposition~\ref{Proposition1}, we design an INR, CoefINR, to indirectly optimize the beamforming function by learning the weighting coefficient function on the user CAPA. The beamforming function is then reconstructed from~\eqref{optimal structure}. Proposition~1 shifts the learning target from a function defined on the BS CAPA to one defined on the user CAPA. Since the user CAPA typically spans a smaller area than the BS CAPA, when applying GL quadrature to approximate the integrals, the integrals defined on $\mathcal{S}_{\mathrm{U}}$ require fewer sampling points compared to those defined on $\mathcal{S}_{\mathrm{B}}$. This reduction in sampling points leads to a lower training complexity. 

Similar to~\eqref{BeaINR}, the CoefINR representing the function $\mathbf{c}(\mathbf{r})$ is modeled as
\begin{equation} \label{INR of C}
\mathbf{c}(\mathbf{r}) = \mathcal{C}_{\theta_{\mathrm{c}}}(\bar{\mathbf{r}}, \mathbf{r}_o),\quad \bar{\mathbf{r}} \in \bar{\mathcal{S}}_\mathrm{U},
\end{equation}
where $\mathcal{C}_{\theta_\mathrm{c}}(\cdot)$ denotes the INR-based weighting coefficient function, parameterized by $\theta_\mathrm{c}$. To train CoefINR, we construct two types of training datasets as in Sec.~\ref{Training Method}. The first uses a fixed set of GL quadrature points across all samples, while the second employs sample-wise randomized sampling based on Owen-scrambled Sobol sequence. These datasets are denoted~as 
\begin{subequations} \label{datasets of CoefINR}
    \begin{align}
    \mathcal{D}_{\mathrm{U,G}} &= \Big\{ \Big( \big\{ \bar{\mathbf{r}}_{m,n} \big\}_{m,n=1}^{M_{\mathrm{U,G}}}, \mathbf{r}_o^{(k)}\Big) \Big\}_{k=1}^{\left|\mathcal{D}_{\mathrm{U,G}}\right|}, \label{GL datasets of CoefINR} \\
    \mathcal{D}_{\mathrm{U,S}} &= \Big\{ \Big( \big\{ \bar{\mathbf{r}}_{i}^{(k)} \big\}_{i=1} ^{M_{\mathrm{U,S}}},\, \mathbf{r}_o^{(k)}\Big) \Big\}_{k=1}^{\left|\mathcal{D}_{\mathrm{U,S}}\right|}, \label{SB datasets of CoefINR} 
    \end{align}
\end{subequations}
where $\bigl\{ \bar{\mathbf{r}}^{(k)}_{i} \bigr\}_{i=1}^{M_{\mathrm{U,S}}}$ are  Owen-scrambled Sobol sampling points independently generated for each sample, $M_\mathrm{U,S}$ is the Sobol sequence length, and $\left|\mathcal{D}_{\mathrm{U,G}}\right|$ and $\left|\mathcal{D}_{\mathrm{U,S}}\right|$ denote the corresponding dataset sizes.  

The training of CoefINR involves two main steps: beamforming function construction and loss evaluation. In the first step, the beamforming function is constructed based on~\eqref{optimal structure}. For instance, for the $k$-th sample in $\mathcal{D}_{\mathrm{U,G}}$, the weighting coefficient function is represented~as
\begin{equation} \label{coefficient vector}
\mathbf{c}^{(k)}(\mathbf{r}_{m,n}^{(k)}) = \mathcal{C}_{\theta_c}(\bar{\mathbf{r}}_{m,n}, \mathbf{r}_o^{(k)}), \quad \bar{\mathbf{r}}_{m,n}\in\bar{\mathcal{S}}_{\mathrm{U}},
\end{equation}
and the beamforming function is approximated as
\begin{align} \label{integral for w}
\textstyle\!\!\mathbf{w}^{(k)}(\mathbf{s}) \!\approx\! \sum\limits_{m=1}^{M_{\mathrm{U,G}}} \sum\limits_{n=1}^{M_{\mathrm{U,G}}} \frac{\xi^{\mathrm{U}}_m \xi^{\mathrm{U}}_n A_{\mathrm{U}}}{4} \,
h(\mathbf{r}^{(k)}_{m,n}, \mathbf{s}) \cdot \mathbf{c}^{(k)}(\mathbf{r}^{(k)}_{m,n}).
\end{align}
In the second step, given $\mathbf{w}^{(k)}(\mathbf{s})$, the loss over $\mathcal{D}_{\mathrm{U,G}}$ and $\mathcal{D}_{\mathrm{U,S}}$ is computed following the same procedure as described in Sec.~\ref{Training Method} after~\eqref{beamforming vector}. The overall training loss is obtained as~\eqref{eq:total_loss}.

\section{Simulation Results} \label{Simulation}
In this section, we evaluate the performance of BeaINR and CoefINR by comparing them with relevant baselines.

\subsection{Simulation Setup}\label{Simulation_setup}
We consider the following simulation setup. The BS and user CAPAs are configured with lengths $L_{\mathrm{B},x}=L_{\mathrm{B},y}=2\,\text{m}$ and $L_{\mathrm{U},x}=L_{\mathrm{U},y}=0.5\,\text{m}$, respectively. The position of user $\mathbf{r}_o=[r_{x}, r_{y}, r_{z}]^{\mathsf{T}}$ is uniformly distributed in a region where $r_{x}, r_{y}$ in $(-5,5)$ m and $r_{z}$ in $(20,30)$~m.  The wavelength is set to $\lambda=0.125\,\text{m}$, and the intrinsic impedance is $\eta=120\,\pi\Omega$. To maximize the multiplexing gain, the number of data streams is set as $N=\min\{M_{\mathrm{B}}, M_{\mathrm{U}}\}$, where $M_{\mathrm{B}} = \big(2\big\lceil \frac{L_{{\mathrm{B}},x}}{\lambda} \big\rceil + 1\big)\big(2\big\lceil \frac{L_{{\mathrm{B}},y}}{\lambda} \big\rceil + 1\big)$ and $M_{\mathrm{U}} = \big(2\big\lceil \frac{L_{{\mathrm{U}},x}}{\lambda} \big\rceil + 1\big)\big(2\big\lceil \frac{L_{{\mathrm{U}},y}}{\lambda} \big\rceil + 1\big)$ according to the analysis in~\cite{Beamforming_Design}. The importance factor in~\eqref{eq:total_loss} is $\alpha=0.1$. The noise variance is set to $\sigma^2=5.6\times10^{-3}\,\text{V}^2$, and the budget of the total and peak transmit current are set to $\mathrm{C}_{\max}=800\,\text{mA}^2$ and $\mathrm{I}_{\max}=200\,\text{mA}^2/\text{m}^2$, respectively.

Two fully connected neural networks are employed as CoefINR and BeaINR.\footnote{From \eqref{BeaINR} and \eqref{INR of C}, the functions learned by BeaINR and CoefINR do not exhibit permutation equivariance or translation invariance properties, and thus GNNs or convolutional neural networks (CNNs) that exploit these properties are not applicable, while FNNs are an appropriate architectural choice for both networks.} The hidden-layer dimensions of both networks are set to $[32, 512, 512, 1024, 512, 512, 32]$, with Tanh activations applied to all hidden layers. 
The input dimension of both BeaINR and CoefINR is 5. For BeaINR, the input consists of the user position coordinates $\mathbf{r}_o=[r_{x}, r_{y}, r_{z}]^{\mathsf{T}}$ and the spatial coordinates $(s_x, s_y)$ of a sampling point $\mathbf{s}=[s_x, s_y, 0]^{\mathsf{T}}$ on the BS CAPA. For CoefINR, the input consists of the same user position coordinate and the spatial coordinates $(\bar{r}_x, \bar{r}_y)$ of a sampling point $\bar{\mathbf{r}}=[\bar{r}_x, \bar{r}_y, 0]^{\mathsf{T}}$ on the user CAPA. The output dimensions of both networks are $2N$ corresponding to the real and imaginary parts of $\mathbf{w}(\mathbf{s})$ or $\mathbf{c}(\mathbf{r})$ for the $N$ data streams. To satisfy the peak current constraint in~\eqref{eq:PeakConstraint}, the output of BeaINR employs the following normalization activation $\mathbf{w}\left( \mathbf{s} \right) =\bar{\mathbf{w}}\left( \mathbf{s} \right) \sqrt{\frac{\mathrm{I}_{\max}}{\left\| \bar{\mathbf{w}}\left( \mathbf{s} \right) \right\| ^2+\mathrm{I}_{\max}}}$, where $\bar{\mathbf{w}}\left( \mathbf{s} \right)\in\mathbb{C}^{1\times N}$ is the unnormalized output, and $\left\| \mathbf{w}\left( \mathbf{s} \right) \right\|\in[0, \mathrm{I}_{\max})$. The output of CoefINR does not apply any activation~function.

For training, we generate 500,000 samples in $\mathcal{D}_{\mathrm{U,G}}$, $\mathcal{D}_{\mathrm{U,S}}$, $\mathcal{D}_{\mathrm{B,G}}$ and  $\mathcal{D}_{\mathrm{B,S}}$. Each sample in $\mathcal{D}_{\mathrm{U,G}}$ contains $M^2_{\mathrm{U,G}}=100$ fixed GL sampling points on $\mathcal{S}_{\mathrm{U}}$, while each sample in $\mathcal{D}_{\mathrm{U,S}}$ includes $M_{\mathrm{U,S}}=M^2_{\mathrm{U,G}}$ randomized Sobol sampling points.\footnote{The quadrature order determines the trade-off between integral-approximation accuracy and training complexity: increasing $M_{\mathrm{U,G}}$ reduces the approximation error but increases computational cost. Our simulations show that the approximation is sufficiently accurate for $M_{\mathrm{U,G}}\geq8$. Thus, we set $M_{\mathrm{U,G}}=10$ to ensure accuracy with moderate computational overhead.} To ensure consistent sampling density across both CAPAs, the numbers of sampling points are scaled to the CAPA areas. Thus, the number of the GL sequence and the Sobol sampling points on $\mathcal{S}_\mathrm{B}$ are set as $M_{\mathrm{B,G}}\!=\!\left\lceil\!\frac{L_{\mathrm{B},x}}{L_{\mathrm{U},x}}\!\right\rceil \!M_{\mathrm{U,G}}$ and $M_{\mathrm{B,S}}\!=\!\left\lceil\!\frac{L^2_{\mathrm{B},x}}{L^2_{\mathrm{U},x}}\!\right\rceil \!M_{\mathrm{U,S}}$, respectively. For testing, we generate another 10,000 samples in $\mathcal{D}_{\mathrm{U,G}}$ and $\mathcal{D}_{\mathrm{B,G}}$. To achieve accurate integral approximations in testing, a higher-resolution GL sampling grid is employed, with $M_{\mathrm{U,G}}=20$ and $M_{\mathrm{B,G}}\!=\!\left\lceil\!\frac{L_{\mathrm{B},x}}{L_{\mathrm{U},x}}\!\right\rceil \!M_{\mathrm{U,G}}$.  
Both models are trained using the Adam optimizer with an initial learning rate of 0.0001, and all simulations are carried out on a computer equipped with an Intel(R) Core(TM) i9-9980XE CPU (3.00 GHz) and an NVIDIA GeForce RTX 2080 Ti GPU.

\subsection{Learning Performance} 
We compare the learning performance of BeaINR and CoefINR against the following numerical baselines:
\begin{itemize}
    \item \textbf{WMMSE}: This method adopts the CoV-based approach~\cite{Beamforming_Design}, where the beamforming function is optimized using an iterative WMMSE algorithm and the involved integrals are approximated by the GL quadrature.

    \item \textbf{Fourier}: As proposed in~\cite{Wavenumber}, the beamforming function is expressed by a finite set of Fourier series, and the coefficients of the Fourier basis functions are optimized.
    
    \item \textbf{SPDA}: Following the discretization strategy in~\cite{Pattern_Division}, the continuous CAPA is discretized into finite elements, reducing the problem to an SPDA formulation, which is solved via singular value decomposition and water-filling.
\end{itemize}  
These baselines do not consider the peak transmit current constraint in~\eqref{eq:PeakConstraint}. To address this issue, the resulting beamforming solutions are uniformly scaled to satisfy the~constraint.

{Fig.~\ref{Performance_comparison_under_different_current_constraints} illustrates the SE achieved under different total current constraint $\mathrm{C}_{\max}$ and peak current constraint $\mathrm{I}_{\max}$. As shown in~\figsubref{Performance_comparison_under_different_current_constraints}{SE_Vs_Pmax}, with a fixed $\mathrm{I}_{\max}=200\text{mA}^2/\text{m}^2$, SE increases with $\mathrm{C}_{\max}$ and then saturates once the system becomes peak-current-limited.
Conversely,~\figsubref{Performance_comparison_under_different_current_constraints}{SE_Vs_Imax} shows that with a fixed $\mathrm{C}_{\max}=1000\mathrm{mA}^2$, SE increases with $\mathrm{I}_{\max}$ until the system becomes total-current-limited. Fig.~\ref{SE_VS_user_Size} demonstrates that SE increases monotonically with the user CAPA area due to the enhanced spatial degrees of freedom. In all cases, BeaINR and CoefINR consistently outperform the Fourier and SPDA methods due to avoiding the discretization loss, and they also achieve comparable SE to that of the WMMSE algorithm.

\begin{figure}[h]
\centering 
\vspace{-0.2cm} 
\subfloat[SE versus total current $\mathrm{C}_{\max}$.]{\includegraphics[width=0.46\linewidth]{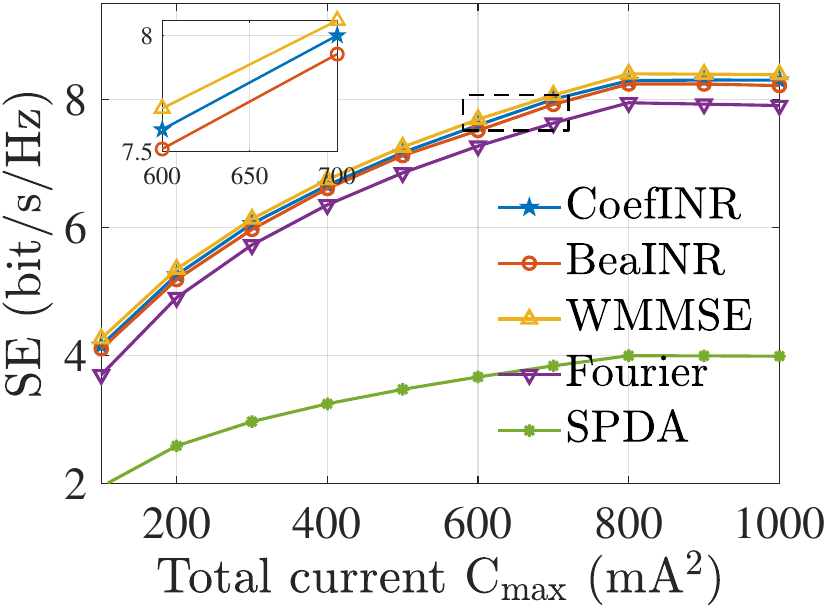} \label{SE_Vs_Pmax}}
\hfill
\subfloat[SE versus peak current $\mathrm{I}_{\max}$.]{\includegraphics[width=0.45\linewidth]{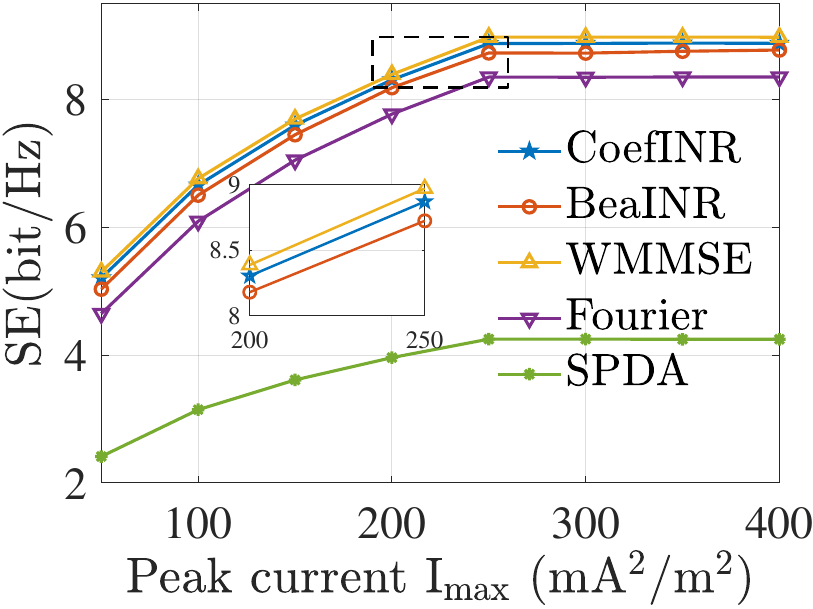}\label{SE_Vs_Imax}  } 
\caption{Performance comparison under different constraints} 
\label{Performance_comparison_under_different_current_constraints} \vspace{-0.1cm} 
\end{figure}

\begin{figure}
\vspace{-0.1cm} 
\centering
\includegraphics[width=0.3\textwidth]{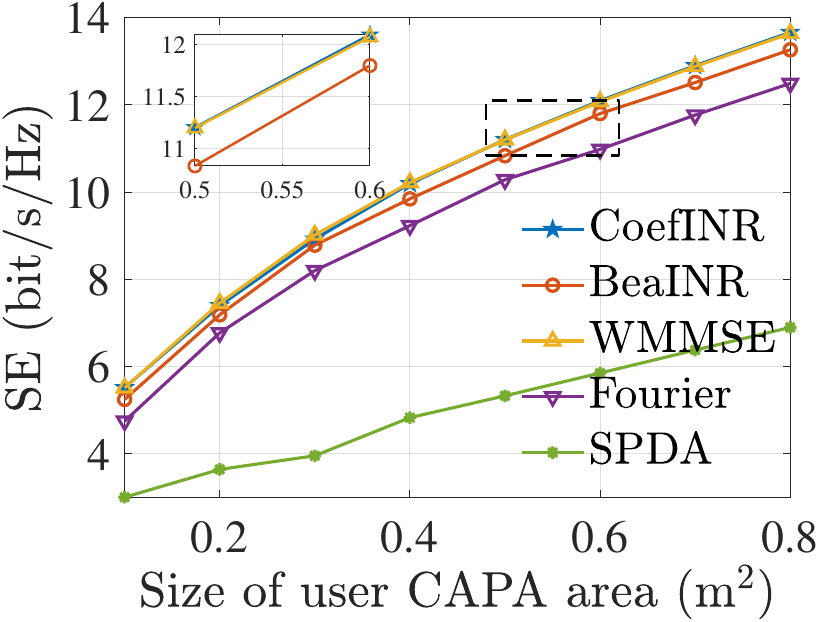}
\caption{SE versus user CAPA size.}  \label{SE_VS_user_Size}
\vspace{-0.4cm}
\end{figure}

\subsection{Generalizability}
In this subsection, we evaluate the generalizability of
BeaINR and CoefINR with respect to CAPA size, carrier frequency, and distance range. 

For the generalization to BS CAPA size, both models are trained with $A_{\mathrm{B}}\!=\!4\,\mathrm{m}^2$ and tested over $A_{\mathrm{B}}\in[2, 6]\,\mathrm{m}^2$. For the generalization to user CAPA size, both models are trained with $A_{\mathrm{U}}\!=\!0.5\,\mathrm{m}^2$ and tested over $A_{\mathrm{U}}\in[0.3, 0.7]\,\mathrm{m}^2$. The performance metric is defined as the ratio of the average SE achieved by BeaINR or CoefINR to that achieved by WMMSE. As shown in Table~\ref{Generalizability}, both networks maintain over 96\% of the WMMSE performance across $A_{\mathrm{B}}$, indicating strong generalizability with respect to the BS CAPA area. Both networks also generalize well to larger user CAPA~sizes.

\begin{table}[h]
\vspace{-0.15cm}
\centering
\captionsetup{font=small}
\caption{CAPA Size Generalizability}
\label{Generalizability}
\begin{tabular}{c|ccccc}
\hline\hline
$A_{\mathrm{B}}$ (m$^2$) & 2 & 3 & 4 & 5 & 6 \\
\hline
BeaINR (\%) & 96.51  & 96.85 & 98.57 & 97.39 & 96.76 \\
CoefINR (\%) & 98.34  & 98.93 & 99.01 & 98.84 & 98.54 \\
\hline
$A_{\mathrm{U}}$ (m$^2$) & 0.3 & 0.4 & 0.5 & 0.6 & 0.7 \\
\hline
BeaINR (\%) & 93.77 & 94.87 & 96.59 & 96.01 & 95.88 \\
CoefINR (\%) & 76.25 & 92.46 & 99.17 & 97.06 & 98.88 \\
\hline\hline
\end{tabular}
\vspace{-0.15cm}
\end{table}

Table~\ref{Frequency_Generalizability} presents the generalization of BeaINR and CoefINR to carrier frequency. Both networks are trained at a fixed carrier frequency $f\!=\!2.4\,\text{GHz}$ and tested over $f\!\in\! [1.8, 3]\,\text{GHz}$. As shown in Table~\ref{Frequency_Generalizability}, BeaINR maintains over 90\% for $f\!\in\! [2.2, 2.6]\,\text{GHz}$, indicating good generalizability in a relatively narrow band. CoefINR generalizes well over the entire tested frequency range. This is because the channel function depends on the wavelength $\lambda$, and thus on the carrier frequency $f$ as shown in~\eqref{Green's function}. By exploiting the optimal solution structure in Proposition~1, CoefINR learns only the weighting coefficient function and uses the frequency-dependent channel function to construct the beamforming function, which leads to improved frequency~generalizability.

\begin{table}[h]
\vspace{-0.2cm}
\captionsetup{font=small}
\centering
\caption{Frequency generalizability}
\label{Frequency_Generalizability}
\setlength{\tabcolsep}{3pt} 
\begin{tabular}{@{}c|*7{c}@{}}
\hline\hline
$f$ (GHz)    & 1.8   & 2     & 2.2   & 2.4   & 2.6   & 2.8   & 3    \\
\hline
BeaINR (\%)  & 83.25  & 89.63 & 95.73 & 97.24 & 94.02 & 87.96 & 80.08 \\
CoefINR (\%) & 96.73  & 98.09 & 99.71 & 100.15 & 99.53 & 98.20 & 96.41 \\
\hline\hline
\end{tabular}
\vspace{-0.2cm}
\end{table}

For the generalization to distance range, both networks are trained with user positions $\mathbf{r}_o=[r_{x}, r_{y},r_{z}]^{\mathsf{T}}$ drawn from the region $r_{x}, r_{y}$ in $(-5,5) \mathrm{m}$  and $r_{z}$ in $(20,30)\mathrm{m}$ and tested under a larger region $r_{x}, r_{y}$ in $(-10,10)\mathrm{m}$ and $r_{z}$ in $(10,40)\mathrm{m}$. The generalized performance achieved by BeaINR and CoefINR is 90.23\% and 89.06\%, respectively.

\subsection{Inference and Training Complexity}
Table~\ref{Complexity} compares the inference time and training complexity of the considered methods. Training complexity is evaluated in terms of sample, time, and space requirements, all measured under the condition of achieving $95\%$ of the SE attained by the WMMSE algorithm. 

As shown in Table~\ref{Complexity}, the learning-based methods yield notably lower inference time compared to numerical baselines. Moreover, CoefINR further reduces training complexity compared to BeaINR, because it reduces the number of sampling points for integral approximations as analyzed in Sec.~\ref{CoefINR Sec}.

\begin{table}[htbp]	
\vspace{-0.1cm}
\captionsetup{font=small}
\centering 
\caption{Inference Time and Training Complexity}	
\begin{tabular}{c|c|c|c|c}
\hline\hline
\multirow{2}{*}{Name} & \multirow{2}{*}{Inference time} & \multicolumn{3}{c}{Training Complexity} \\
\cline{3-5}
                     &                                  & Sample  & Time      & Space \\ \hline             
\textbf{BeaINR}      & 0.062 s                         & 25 K       & 36.2 h   & 4.72 M    \\ 
\cline{1-5}
\textbf{CoefINR}      & 0.043 s                          & 20 K     & 3.68 h     & 2.63 M   \\ 
\cline{1-5}
\textbf{WMMSE}     & 0.378 s                            & \diagbox{}{} & \diagbox{}{}  & \diagbox{}{}  \\ 
\cline{1-5}
\textbf{Fourier}     & 10.88 s                            & \diagbox{}{} & \diagbox{}{}  & \diagbox{}{}  \\ 
\cline{1-5}
\textbf{SPDA}     & 9.590 s                            & \diagbox{}{} & \diagbox{}{}  & \diagbox{}{}  \\ 
\hline\hline
\end{tabular}
\begin{minipage}{0.95\linewidth}
\quad\footnotesize \textit{Note:} “K” and “M” represent thousand and million, respectively.
\end{minipage}
\label{Complexity}
\vspace{-0.1cm}
\end{table}

\subsection{Quantization Effects}
Due to cost and hardware limitations, the beamforming weights in certain CAPA systems are quantized with finite resolution for both amplitude and phase. To quantify this impact, we quantize the continuous beamforming weights learned by BeaINR and CoefINR with different bits during testing. As shown in Fig.~\ref{Quantization_performance}, coarse quantization in either amplitude or phase noticeably degrades SE, while the loss diminishes rapidly as the quantization resolution increases. In addition, the comparison between \figsubref{Quantization_performance}{AmpBit200mA} and \figsubref{Quantization_performance}{PhaseBit200mA} shows that phase quantization is more detrimental than amplitude~quantization.

\begin{figure}[htbp] \centering 
\vspace{-0.1cm}
\subfloat[Amplitude quantization.]{\includegraphics[width=0.46\linewidth]{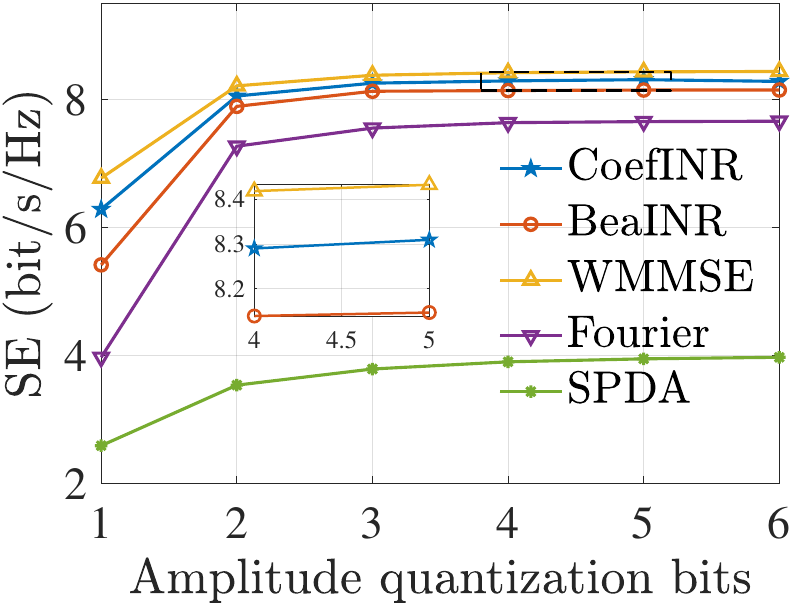}\label{AmpBit200mA} }
\hfill
\subfloat[Phase quantization.]{\includegraphics[width=0.45\linewidth]{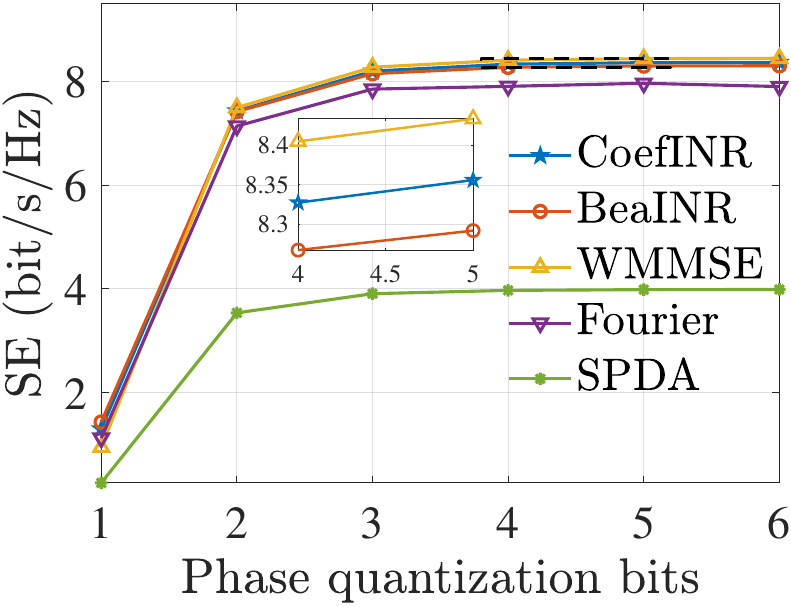} \label{PhaseBit200mA} } 
\caption{SE versus quantization bits.} \label{Quantization_performance} 
\vspace{-0.1cm}
\end{figure}

\section{Conclusions}
This paper proposed INR-based methods for learning the beamforming of CAPA systems, where both the BS and the user are equipped with CAPAs. Two networks, BeaINR and CoefINR, were designed to learn the beamforming function, either directly or via the optimal beamforming structure. Simulation results show that the proposed methods achieve comparable or higher SE than baselines with significantly reduced inference latency. CoefINR improves training efficiency and frequency generalizability over BeaINR by explicitly exploiting the optimal beamforming structure.

In future work, extending the proposed approach to multiuser multi-CAPA systems and developing robust beamforming methods against channel estimation errors are important research directions.

\bibliographystyle{IEEEtran}
\bibliography{main}
\end{document}